\documentclass[hidelinks]{article}
\usepackage[T1]{fontenc}
\usepackage[a4paper]{geometry}
\usepackage{listings}
\usepackage{multicol}
\usepackage[ruled,vlined]{algorithm2e}
\usepackage{algcompatible}
\usepackage{babel}
\usepackage{amsmath}
\usepackage{amssymb}
\newcommand{\specialcell}[2][c]{%
	\begin{tabular}[#1]{@{}l@{}}#2\end{tabular}}

\usepackage[]{authblk}
\usepackage{mathtools}
\usepackage{setspace} 
\usepackage{hyperref} 

\usepackage[none]{hyphenat}

\usepackage[round]{natbib} 

\usepackage{float}
\begin{document}
	
\title{Can an AI agent hit a moving target? 
	\thanks{I am grateful for the advice and constant support of Prof. Roger Farmer and Prof. Herakles Polemarchakis.  I appreciate the financial support from Warwick University and the many conversations with colleagues at the Rebuilding Macroeconomics project. I am grateful for the helpful discussions and guidance from colleagues at the Bank of England, which also inspired the choice of the DDPG algorithm I used in this exercise. I thank participants at the 2021 Money Macro Finance annual conference, the Society for Computational Economics 28th International Conference, the CES 2022 annual conference, and the BSE PhD Workshop on Expectations in Macroeconomics. All remaining errors are mine.}
}

\author{\large Rui (Aruhan) Shi \thanks{PhD candidate at the Department of Economics,
		University of Warwick. Email: \href{mailto:a.shi@warwick.ac.uk}{a.shi@warwick.ac.uk}.}
		
}


\date{\today\\ \href{https://warwick.ac.uk/fac/soc/economics/staff/rshi/jmp.pdf}{Click here for the latest version.}
} 


\maketitle

\thispagestyle{empty}

\begin{abstract}
I model the belief formation and decision making processes of economic agents during a monetary policy regime change (an acceleration in the money supply) with a deep reinforcement learning algorithm in the AI literature. I show that when the money supply accelerates, the learning agents only adjust their actions, which include consumption and demand for real balance, after gathering learning experience for many periods. This delayed adjustments leads to low returns during transition periods. Once they start adjusting to the new environment, their welfare improves. Their changes in beliefs and actions lead to temporary inflation volatility. I also show that, 1. the AI agents who explores their environment more adapt to the policy regime change quicker, which leads to welfare improvements and less inflation volatility, and 2. the AI agents who have experienced a structural change adjust their beliefs and behaviours quicker than an inexperienced learning agent. 

\end{abstract}


\newpage

\section{Introduction}
\label{section_intro}




Inflation is rising around the world, adding further disturbance to an already uncertain global economy. It is more important than ever for policy makers to understand transition dynamics. In this paper, I adopt a deep reinforcement learning (DRL) algorithm from the artificial intelligence (AI) literature to model economic agents' learning processes, and present transition dynamics of an economy with a change in the monetary policy regime in a dynamic stochastic general equilibrium (DSGE) model. I model agents' belief formation and decision-making processes during an acceleration of money supply. Through simulation experiments, I show how agents pick up such an acceleration, how their beliefs and decisions adapt to this change, and how their actions change the aggregate economic dynamics. I also show that agents have heterogeneous responses to the same macro environment.


DRL algorithms produce many successes in fields outside of economics. For example, WaveNet is used for Google Assistant and the latest android devices for voice recognition \citep{Wavenet}; Deepmind AI reduces the Google data centre cooling bill by 40\% \citep{Googlecooling}.  

I apply a DRL algorithm in a general equilibrium model for three reasons: 
\begin{itemize}
    \item a DRL learning agent picks up structural changes, such as an acceleration of the money supply, and their beliefs and actions impact the aggregate transition dynamics of the economy;
    \item learning agents have heterogeneous beliefs and actions owing to the behavioural parameter exploration and a memory component that stores past experience;
    \item the memory component provides a theoretical framework to model experience-based learning.
\end{itemize} 

In this setting, the agent first interacts with an environment by taking random actions and receiving rewards. This process of learning based on rewards received is inspired by learning through trial and error in the psychology of animal learning \citep{SB2018}. The randomness of the action is linked to how exploratory the agent is in terms of trying different options in an action space. This can be viewed as a hardwired trait in the agent. The agent has to exploit what it has already experienced to obtain rewards, but it also has to explore to make better selections in the future. 

Different levels of exploration also lead to different learning behaviours. This means that in the same environment and facing the same state variables, learning agents behave differently based on their willingness to explore. With a high level of exploration, the agent experiences a wide range of possibilities. This ensures the agent has sufficient experience to learn and adapt to environmental changes. 

The agent uses samples from past experience to update beliefs. Beliefs are embedded in the value function of the agent. When the agent's beliefs change, the policy function adjusts and generates different decisions. Both the decision-making strategy and the value function are approximated by randomly initialised artificial neural networks (ANNs). ANNs can be viewed as flexible function approximators. Functional forms (e.g., linear or quadratic) of the policy and the value functions need not to be known before learning. This implies that the agent is not learning about a particular model parameter or group of model parameters, but is learning how to make decisions based on past experience, and adjust subjective beliefs about the world. This adjustment of beliefs does not follow Bayes' rule.  

I set up the monetary authority to increase the inflation target and accelerate the nominal money supply. This setup is inspired by the early literature on the accelerationist controversy. An accelerationist or backward-looking Phillips curve with adaptive expectation permits a trade-off between inflation and unemployment. It was argued that a government could exploit such an opportunity to maintain a low rate of unemployment by accelerating the money supply process. 

As the agent in this AI algorithm learns from past experience, it is a form of adaptive expectation. Through this exercise, I show how AI agents are "smart" enough to adjust to the new policy regime in a general equilibrium model, and how past experiences impact their behaviours during a regime shift. This is important for policy experiments in a realistic learning environment with many (heterogeneous) artificial agents living in a general equilibrium framework. 



 
 The rest of the paper is organized as follows. Section \ref{lit} provides a literature review. Section \ref{model} describes the economic model.  Section \ref{AI} introduces and discusses the AI framework of bounded rationality. Section \ref{experiments} and \ref{results} discuss the experiments conducted, the results, and the discussions. The last section concludes.

\vspace{1.0cm}

\section{Related Literature}
\label{lit}

This paper contributes to the literature on modelling agents' expectation formation processes. Keynes invoked the importance of agents' expectations' when he showed how expectations determine output and employment \citep{Keynes:1936}. Two decades later, \cite{Cagan1956} and \cite{Friedman1957} formalised the idea of adaptive expectations. In combination with the Phillips curve, their proposal generated a large debate about whether and how a government could exploit a possible negative relationship between inflation and unemployment. However, it was criticised for the assumption that an agent would make an inflation forecast that was the same as past period inflation. 

The alternative, which revolutionised macroeconomics, was the rational expectations hypothesis \citep{LUCAS1972103, LUCAS197619, Sargent1971, Sargent19733}. This assumes that agents make model-consistent beliefs, and they understand the economy. In other words, the agents go from very naive (adaptive expectation) to very smart. It has many advantages, one of which is its usefulness for thinking about policy experiments in a relatively stationary environment. 

It also has some disadvantages, one of which is that it does not offer convincing dynamics of inflation in response to shocks. Many techniques have been proposed to take the middle ground between rational expectations, in which agents are too smart,  and adaptive expectations, in which agents are too naive. These can be broadly classified into two groups.


One group explores the implications of information rigidities, and this includes sticky information \citep{MankiwReis2002, Balletal2005}, noisy information \citep{Woodford2001} and rational inattention \citep{SIMS2003665}. The main focus is that agents are constrained in obtaining or processing information, and thus only use a portion of the full information to make `optimal' decisions, i.e., still hold model-consistent beliefs but with less than full information. Similar to this literature, I argue that agents are constrained in the amount of information they can collect and process at any given time. However, the difference is that the agents do not form model-consistent beliefs. They form decision-making strategies based on their own experiences.


The other group focuses on bounded rationality \citep{Sargent1993} and adaptive learning \citep{EvansHonkapohja1999}. \cite{Schorfheide2005}, \cite{OzdenWouter2021}, \cite{Airaudo2021} look at combining adaptive learning with Markov switching specifications to model learning agents with policy regime changes. The agents in the adaptive learning literature are believed to be as smart as econometricians, and learn about model parameters by running a regression with past data or applying Bayesian updating. Using AI algorithms to model decision making processes is also a form of adaptive learning, since the learning is based on past experience. However, the AI learning agents learn from their own experiences in the environment by making exploratory actions. 
 
 


This paper is similar to recent literature that uses ANNs to model economies. \cite{ashwin2021unattractiveness} study the stability properties of multiple equilibria with learning agents. Their agents learn with ANNs. Similarly, \cite{kuriksha2021economy} models economic agents with deep ANNs in a macro-financial environment. My learning agents generate their own experiences by interacting with an environment. This differs from \cite{ashwin2021unattractiveness} and \cite{kuriksha2021economy} in using only deep learning methods (deep ANNs).

I use an algorithm from the fast-moving AI literature, which belongs to deep reinforcement learning (DRL) algorithms. For example, machines trained with the deep Q network algorithm \citep{mnih-atari-2013} are capable of human-level performance on many Atari video games using unprocessed pixels for input. However, the deep Q network algorithm can only handle discrete action spaces. The deep deterministic policy gradient algorithm (DDPG) from \cite{lillicrap2015drl} is often used in learning settings with continuous action spaces. This algorithm is more applicable in an economic model, given that economic decision-making processes often involve continuous action spaces.

The application of DRL algorithms in macroeconomic models represents a new branch of research. In a companion paper, \cite{Shi2021learning} adopts a DRL algorithm in a stochastic growth model. She shows that an AI agent learns from no information on the economic structure nor its preference, and it has the ability to adapt to transitory and permanent income shocks. Most recent literature look into using DRL algorithms as a solution method. \cite{Shi2021deep} apply a DRL algorithm in a model with different monetary and fiscal policy regimes, and show evidence that a DRL agent can locally converge to all equilibria in the model. \cite{Hinterlangand2021} use a DRL algorithm to solve for the optimal policy response function. \cite{https://doi.org/10.48550/arxiv.2103.16977} and \cite{https://doi.org/10.48550/arxiv.2201.01163} look into using a DRL algorithm to solve for multi-agent macroeconomic models.

\section{An Economic Model with the Rational Expectations Assumption}
\label{model}

In this section, I present a general equilibrium model with a representative agent that follows the rational expectations assumption. The model is similar to \cite{Sims1994}.

\subsection{A Representative Household}

A representative household aims to maximise its lifetime utility.

\begin{equation}
	E_0\sum_{t=0}^{\infty} \beta^t u(c_t),
\end{equation}

where $\beta \in (0,1)$,

subject to a nominal budget constraint,








\begin{equation}
	P_t c_t (1 - f(v_t)) + P_t s_t  + M_t \leq P_t y_t +  M_{t-1} + P_t r_{t-1} s_{t-1} - P_t \tau_t
	\label{NBC}
\end{equation}

where $P_t$ is the price level at period $t$, $c_t$ denotes the consumption at $t$, $M_t$ is nominal money balance, $s_{t-1}$ is the saved stock of goods that a household enters period $t$ with, and there is a storage technology that pays out an interest rate $r_{t-1}$. $y_t$ is the endowment or income of the agent. $\tau_t$ is the government transfer at $t$. Velocity is defined as $v_t \equiv \frac{c_t}{m_t}$, and $f(v)$ is the transaction cost function (per consumption unit). It is assumed to take the form $f(v_t) = v_t$. In real terms, the budget constraint is as follows,

\begin{equation}
	c_t (1 - f(v_t))+ s_{t} + m_{t} = y_t + m_{t-1} \frac{P_{t-1}}{P_t} + r_{t-1} s_{t-1} - \tau_t 
	\label{bcc}
\end{equation} 

where real balance is defined as $m_t \equiv \frac{M_t}{P_t}$. Inflation is defined as $\pi_t \equiv \frac{P_t}{P_{t-1}}$. Inflation has an effect on the real economy through the transaction cost function. Transaction costs take away real resources. 

The endowment depends on a constant $\bar{y}$ and an exogenous process $\epsilon_t^y$,
\begin{equation}
	y_t = \bar{y} + \epsilon_t^y
	\label{endowment}
\end{equation}

The interest rate is determined exogenously, and depends on a constant $\bar{r}$ and an exogenous process  $\epsilon_t^r$,
\begin{equation}
	r_t = \bar{r} + \epsilon_t^r
	\label{interest}
\end{equation}

\subsection{Government: Fiscal and Monetary Policies}

The government supplies money based on a money growth rule.

\begin{equation}
	M_t = \delta^MM_{t-1}
\end{equation}

$\delta^M$ is a policy variable that determines the speed of money supply.

The government's budget constraint is

\begin{equation}
	g_t - \tau_t = m_t - \frac{m_{t-1}}{\pi_{t}}.
	\label{g_bc}
\end{equation}

The government sets initial nominal money supply $M_0$ and the policy variable $\delta^M$. It also determines the taxation $\tau_t$. Government spending $g_t$ reacts in response to changes in inflation.

\section{An AI Learning Framework with a DRL Algorithm}
\label{AI}

I introduce the DDPG algorithm\footnote{For a comprehensive review of reinforcement learning, please see \cite{SB2018}.} and show how to use it to model economic agents' decision-making process in the model specified in section \ref{model}.

\subsection{An AI Learning Framework: the Actor-Critic Model}
\begin{figure}[H]	
	\caption{The agent-environment interaction in a reinforcement learning setting}
	\centerline{\includegraphics[width=12cm,height=5cm]{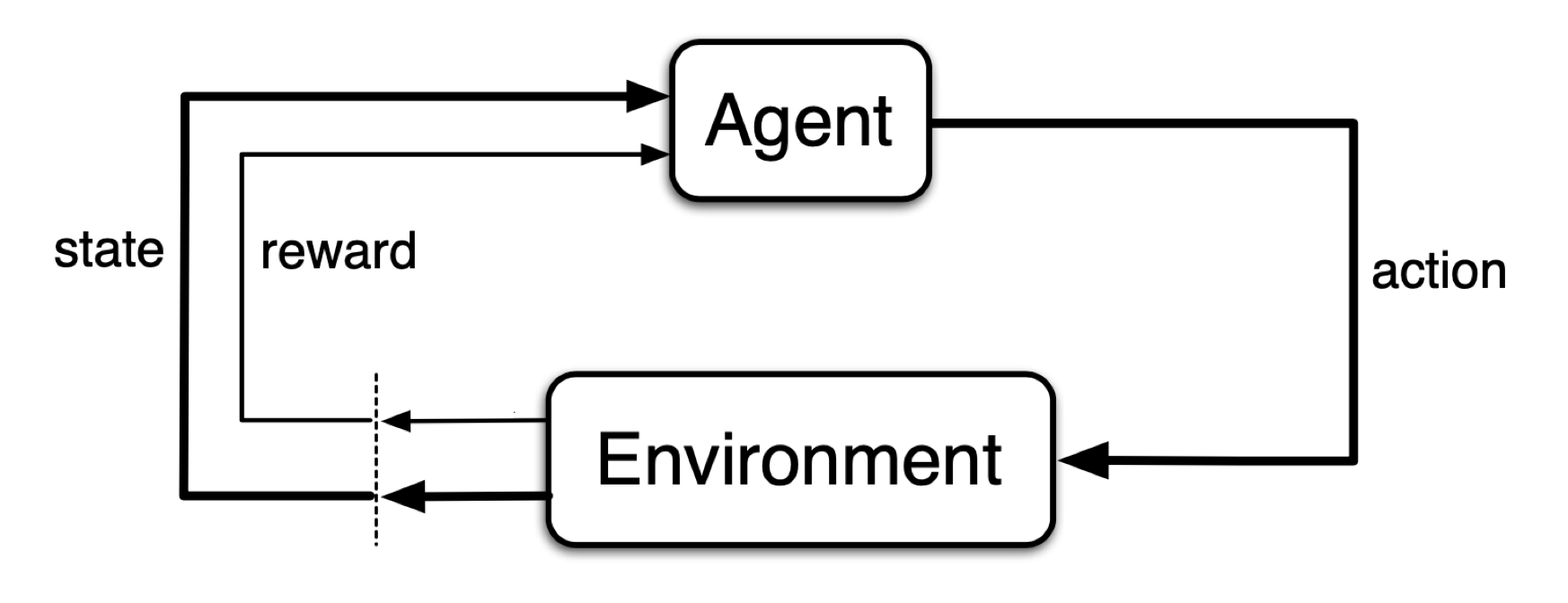}}
	\label{MDP}
	Source: \cite{SB2018}
\end{figure}

The DRL algorithm adopted here was first introduced by \cite{lillicrap2015drl}, namely deep deterministic policy gradient (DDPG). \cite{Shi2021deep} also adopt the DDPG algorithm in their study of learnability of rational expectations equilibrium in different policy regimes. This algorithm mainly follows the actor-critic model of reinforcement learning, and it uses the formal framework of a Markov Decision Process (MDP) to define the interactions between a learning agent and its environment in terms of states, actions, and rewards (Figure \ref{MDP}).

\vspace{1.5cm}

\vspace{1.5cm}

State $S$ is a random variable from a state space, which is a bounded and compact set,\footnote{The latest research on reinforcement learning also investigates the setting with unbounded state space, e.g., \cite{shah2020stable}.} i.e., $S \in \mathcal{S}$. The agent takes an action $A$, which belongs to an action space $\mathcal{A}$,  $A\in\mathcal{A}$. The state evolves through time following a probability function, $p: \mathcal{S} \times \mathcal{S} \times \mathcal{A} \rightarrow [0,1]$, which is defined as

\begin{equation}
	p(S'|S, A) \equiv Pr \{S_t = S' | S_{t-1} = S, A_{t-1} = A \}.
\end{equation}

It shows the probability of a random variable state $S'$ occurring at time $t$, given the preceding values of state, $S$, and action, $A$.

Reward is a random variable and can be generated from a reward function, $R: \mathcal{S} \times \mathcal{A} \rightarrow \mathcal{R}$.

The return from a state is defined as the sum of discounted future rewards,

\begin{equation}
	G_t \equiv R_{t} + R_{t+1} + ... = \sum_{k=0}^{\infty} \beta^k R_{t+k},
\end{equation}
where $\beta$ is the discount factor.


A RL learning agent's behaviours follow a policy function, also known as the actor network. The policy function can be both stochastic and deterministic. A stochastic policy maps states to probabilities of selecting each possible action. A deterministic policy, which is used in this paper, maps a state from a state space to an action from an action space, and it is denoted by $\mu: \mathcal{S} \rightarrow \mathcal{A}$.

A value function\footnote{In reinforcement learning literature, two types of value functions are defined: action-value function and value function. To not complicate the matter, I use value function as action-value function throughout this paper.}, also known as the critic network, shows the `expected' returns of taking an action in a state and thereafter following policy $\mu$. Expectations here are subjective beliefs that depend on learning agents' past experiences. A value function is defined as,

\begin{equation}
	Q^{\mu}(S,A) \equiv E^\mu [G_t|S_t = S, A_t = A],
\end{equation}

where $Q^\mu$ means the action value function follows policy $\mu$, and $E^\mu$ reflects that it's a subjective belief that depends on a policy $\mu$ that is formed by past experience.\footnote{This forms of notation, e.g., $E^\mu$, largely follows The Handbook for Reinforcement Learning by \cite{SB2018}.}

Many approaches in reinforcement learning make use of the recursive relationship known as the Bellman equation,

\begin{equation}
	Q^{\mu}(S,A) = R(S, A) + \beta E^\mu Q(S', A'),
\end{equation}

where $A' = \mu(S')$.

RL methods focus on how the learning agents' policy and value functions change as a result of their experience. These changes can be a functional form change or parameter value updates. The DDPG algorithm uses ANNs to approximate policy and value functions: the actor network is denoted as $\mu(S|\theta^\mu)$, where $\theta^\mu$ represents parameters of the ANNs; the critic network is denoted as $Q(S, A|\theta^Q)$, and $\theta^Q$ is its parameters. $\theta^\mu$ and $\theta^Q$ are updated during learning, and can be viewed as the coefficients of two functions and the probabilities involved in making subjective expectations. Two ANNs are updated with respect to each other. In the following paragraphs, I highlight key elements on how the actor and critic networks are updated. The full algorithm is presented in Section \ref{fullalgo}.

The goal of RL learning agents is to continuously update their subjective beliefs about the world based on experience, and to form a decision-making strategy (approximated by the actor network) that produces the highest discounted future returns (approximated by the critic network). The actor network is updated with the goal of maximising the corresponding critic network. In other words, the actor network is updated based on what the agent believes, at that time, to be a strategy that produces high `expected' returns. The critic networks evolves over time. What the agent follows as the critic network at period $t$ is different from what it is at $t+1$. Expectations are the learning agents' subjective beliefs that are formed from past experience. 

The critic network is updated with the goal of minimising a TD error (a temporal difference error).\footnote{The full algorithm is in the next section.} The TD error follows the form,

\begin{equation}
	\label{TD error}
	 T - Q(S, A|\theta^Q),
\end{equation}
where $T$ is called a TD target, and it adds the reward given a state-action pair to the discounted values of the next state and action, i.e.,

\begin{equation}
	T = R(S, A) + \beta Q(S',A'|\theta^Q)
\end{equation}

and the next period action $A'$ is assumed to follow the actor network $\mu(S|\theta^\mu)$ at that time.\footnote{It need not be the same as the true policy.}
\begin{equation}
	A' = \mu(S'|\theta^\mu).
\end{equation}

Intuitively, TD targets represent the best possible returns learning agents receive following a state-action pair and their subjective beliefs. 

Neural science research shows that the dopamine neuron firing rates in the brain resemble the TD error sequence during learning \citep{Botvinick2019}. This motivates research in neural science to model decision-making in connection with RL algorithms.

\subsubsection{Exploration}

Explorations play a crucial rule because learning agents that make exploratory actions collect a wide range of information. An exploratory policy is defined as,

\begin{equation}
	\label{OU}
	\mu'(S_t) = \mu(S_t|\theta^\mu) + \mathcal{N}_t.
\end{equation}

This shows that the final action the agent takes, i.e., what $\mu'(S_t)$ generates, depends on the actor network  $\mu(S_t|\theta^\mu)$, and a random variable sampled from a noise process $\mathcal{N}_t$. Following \cite{lillicrap2015drl}, $\mathcal{N}_t$ is sampled from a discretised Ornstein-Uhlenbeck (OU) process.\footnote{There is a strain of literature in computer science that solely focuses on different exploration strategies to achieve the best performance for a given task. It is out of the scope of the current exercise, and not discussed here.}  

This exploratory policy produces random actions. The randomness decreases over time (by design) but it never disappears. The implication is that in a stationary environment, the agents learn their policy functions, and the learnt functions can be similar to the true policy function but may never be identical. However, in a non-stationary environment, explorations allow the policy network to adjust and be flexible to changes in the environment. 

The exploration strategy implies that the policy function can converge to a close region of the rational expectation solution (if it exists), but will not be identical to it. In an environment with structural breaks or regime changes, this exploratory policy allows the learning agent to adjust its expectations and adapt its policy to a new regime.

I focus on the learning process and the transition dynamics using the DDPG algorithm. Most RL algorithms also aim to achieve similar results as dynamic programming but with less computation and without assuming a perfect model of the environment \citep{SB2018}. 


\subsection{Connecting to the DSGE Model}

To implement this expectations formation framework in a DSGE model, I first translate the economic model into components of a MDP, which are presented in Table \ref{RL1}.

\begin{table}[H]
	\centering	
	\caption{RL components and the economic environment}	
	\label{RL1}	
	\begin{tabular}{@{}lll@{}}\\ \hline \hline				
		\textbf{Terminologies} & \textbf{Description} & \textbf{\specialcell{Representation in the\\ economic environment}} \\ \hline		
		\vspace{0.5cm}		
		\textbf{State, $S_t$}  & \specialcell{A random variable from a state space, \\ $S_t \in \mathcal{S}$}  & $S_t = \{ y_t, r_{t-1}, \tau_t, \pi_{t-1}, m_{t-1}, s_{t-1}\}$\\
		\vspace{0.5cm}		
		\textbf{Actions, $A_t$}  & \specialcell{A random variable from an action space, \\ $A_t \in \mathcal{A}$} & $A_t = \{\lambda^s_t, \lambda^m_{t}\}$ \\	
		\vspace{0.5cm}		
		\textbf{Rewards, $R_t$} & A function of state and action  & $R_t = ln(c_t)$ \\	
		\vspace{0.5cm}		
		\textbf{\specialcell{Policy function,\\ $\mu(S|\theta^\mu)$}} & \specialcell{A mapping from state to action, \\ $\mu: \mathcal{S} \rightarrow \mathcal{A}$} & \specialcell{Approximated by a neural network,\\ i.e., actor network; \\ parameterised by $\theta^\mu$ \\to be updated during learning}  \\
		\vspace{0.5cm}
		\textbf{\specialcell{Value function, \\$Q(S,A|\theta^Q)$}} &\specialcell{The `expected' (subjective belief) \\return from taking an action in a state}  &\specialcell{Approximated by a neural network,\\ i.e., critic network; \\parameterised by $\theta^Q$ \\to be updated during learning}\\ \hline \hline
	\end{tabular}
	
\end{table}

\subsection{Full Algorithm and Sequence of Events}
\label{fullalgo}

The full algorithm\footnote{The use of the DDPG algorithm comes from discussions with colleagues at the Bank of England.} consists of three main steps: initialisation, interaction, and learning.
\vspace{1cm}

Step I: Initialisation

\begin{itemize}
	\item In a given environment, design a state space $\mathcal{S}$, a continuous bounded and compact set for random variables specified in Table \ref{RL1}; design an action space $\mathcal{A}$, a continuous bounded and compact set for the action (random) variables. 
	
	\item Set up two deep ANNs: an actor network $\mu(S|\theta^\mu)$ takes the argument of a state from the state space and generates an action within the action space; a critic network $Q(S,A|\theta^Q)$ takes the argument of a realised state-action pair and generates a value. Setting up two ANNs involves determining the input and output dimensions, the specific architectures, the number of layers, the number of nodes per layer, and how nodes are connected. 
	
	\item $\theta^\mu$ represents the parameters of the actor network, and $\theta^Q$ represents the parameters of the critic network.
	
	\item Define a replay buffer $\mathcal{B}$ (called transitions in the DRL literature), which is a memory that stores information that is collected by a DRL agent during the agent-environment interactive process. A transition is characterised by a sequence of variables $(S_t, A_t, R_t, S_{t+1})$.
	
	\item Define a length of $N$, which is the size of a mini-batch. A mini-batch refers to a sample drawn from the memory, $\mathcal{B}$. 
	
	\item Define the total number of episodes $E$ and simulation periods per episode. The higher the episodes, the longer the learning periods.\footnote{In the DRL literature, the AI agent is usually set to learn a particular task or an Atari game. An episode thus means re-starting the game or task, and it ends with a terminal state (i.e., the end result of a game). In an economic environment, however, a clear terminal state can be difficult to specify. Therefore, the concept of episodes only correlates to how long an agent has been learning.}
	
\end{itemize}

For each episode, define the initial state,\footnote{Initial state variables can also be randomly drawn from the state space.} and loop Steps II and III.\\

Step II: The AI agent starts to interact with its environment. This step involves how agents' actions are chosen and how their actions impact the aggregate economy. 

\begin{itemize}
	\item Assume $\tau_t = 0.5$, $\delta^M = 1.02$ (or $1.1$, depends on the regime) and other variables that are known to the agent at period $t$: $y_t, r_{t-1}, \tau_t, \pi_{t-1}, m_{t-1}, s_{t-1}$.
	\item The agent selects a random (based on the randomly initialised actor network) of action variables, $A_t = \mu(S_t|\theta^\mu) + \mathcal{N}_t$, and $A_t$ contains $\lambda^s_t$ and $\lambda^m_t$, and $\mathcal{N}_t$ is a noise attached to the action to ensure exploration, which is sampled from an AR(1) process.
	
	\item The demand for real balance is $m_{t} = (1+\lambda^m_t) m_{t-1}$.
	\item Given the exogenous (to the AI agent) nominal money supply from the government, $M_{t} = \delta^M M_{t-1}$, with aggregate money demand equal to aggregate money supply, price level or inflation can be derived as $\pi_t =\frac{M_{t}}{M_{t-1}}\frac{m_{t-1}}{m_{t}}= \delta^M \frac{m_{t-1}}{m_{t}}$. In a one-agent case, aggregate money demand at period $t$ is $m_{t}$.
	
	\item The amount stored is $s_{t} = \lambda^s_{t} (y_t + \frac{m_{t-1} }{\pi_t} + r_{t-1} s_{t-1 } - \tau - m_{t})$.
	\item $c_t$ is reached from the budget constraint equation \ref{bcc}.
	\item The new state variables are $S_{t+1} = \{y_{t+1}, r_{t}, \tau_{t+1}, \pi_{t}, m_{t}, s_{t}\}$, where $y_{t+1} = \bar{y} + \epsilon^y_{t+1}$ and $r_{t} = \bar{r} + \epsilon^r_{t}$, in which $y_{t+1} = \bar{y} + \epsilon^y_{t+1}$, $\epsilon^y_{t+1}$ is sampled from a distribution $N(0,0.1)$, $\bar{y} = 1$, and $r_{t} = \bar{r} + \epsilon^r_{t}$, $\epsilon^r_{t}$ is sampled from a distribution $N(0,0.1)$, $\bar{r} = 1$.
	\item The reward the agent receives is, $R_t = u(c_t)$.
	\item Store a transition $(S_t, A_t, R_t, S_{t+1})$ in the memory $\mathcal{B}$.

\end{itemize}

Step III: Training the AI agent (when the AI agent starts to learn) for period $N \leq t \leq T $.

\begin{itemize}
	\item Sample a random mini-batch of N transitions $(S_i, A_i, R_i, S_{i+1})$ from the memory $\mathcal{B}$.
	\item Calculate the TD-target values $T_{i}$ for each transition $i \in N$ following
	\begin{equation}
	T_{i} = R_i + \beta Q^{\mu}(S_{i+1},\mu(S_{i+1}|\theta^{\mu})|\theta^{Q})
	\end{equation}
	where $Q^{\mu}(S_{i+1},\mu(S_{i+1}|\theta^{\mu})|\theta^{Q})$ is a prediction made by the critic network with state-action pair $(S_{i+1},\mu(S_{i+1}|\theta^{\mu}))$, and $\mu(S_{i+1}|\theta^{\mu})$ is a prediction made by the actor network with input $S_{i+1}$.
	
%
	\item Obtain $Q(S_i, A_i|\theta^Q)$ from the critic network with input state-action pair $(S_i, A_i)$
	
	
	\item Calculate the average loss for this sample of $N$ transitions
	\begin{equation}
	L = \frac{1}{N}\sum_i\big(T_i-Q(S_i,A_i|\theta^Q)\big)^2
	\end{equation}
	\item Update the critic network with the objective of minimising the loss function $L$.\footnote{This involves applying backpropagation and gradient descent procedures.} 
	
	\item For the policy function, i.e., the actor network, the objective is to maximise the value function predictions. Define the objective function as,
	
	\begin{equation}
	J(\theta^{\mu}) = Q^\mu(S_i, \mu(S_i|\theta^{\mu})|\theta^Q).
	\end{equation}
	
	\item Maximising this objective function is equivalent to minimising $-J(\theta^\mu)$. Update the actor network parameters $\theta^\mu$ with the objective of minimising $-J(\theta^\mu)$.\footnote{Similar to the critic network, the specific steps of updating ANN's parameters by minimising an objective function involve backpropagation and gradient descent.}
	
\end{itemize}

\section{Experiments} 
\label{experiments}
In my experiment I follow the example from the early literature on the accelerationist controversy. I set up an environment with an inflation target of 2\%. I then shift the inflation target to 10\%.\footnote{I do not dive into the reason for this change or the probability of its occurrence. Results also available for a regime change from a constant money supply to an accelerating money supply. } The AI learning agents are born in the environment with a 2\% inflation target. They learn in this environment until the inflation target changes, and this change is unknown to the agents. The aim is to observe their behaviours in response to this unforeseen change, and how the economy transitions. 

This methodological setup and experiment do not contemplate the role of central bank communications.\footnote{It is possible to include the channel of central bank communications, which is left for future research.} Learning and belief updating only happen through learning agents' direct experience, which is consistent with empirical evidence on experience based belief formation.


 \begin{table}[H]
 	\centering	
 	\caption{Main Parameters and Steady State Values under Two Policy Targets}	
 	\label{2target}	
 	\begin{tabular}{@{}lll@{}}\\ \hline \hline				
 		& \textbf{Target I} & \textbf{Target II} \\  \hline		
 		\vspace{0.5cm}		
 		\textbf{Inflation Target, implied by $\delta^M$} &$1.02$ & $1.1$ \\
		\vspace{0.5cm}		
 		\textbf{Inflation, $\bar{\pi}$} &$1.02$ & $1.1$ \\
 		\vspace{0.5cm}		
 		\textbf{Velocity}, $\bar{v}$& $0.17$& $0.32$ \\	
 		\vspace{0.5cm}		
 		\textbf{Discount Factor}, $\beta$& $0.99$& $0.99$ \\  
 		 		\vspace{0.5cm}		
 		\textbf{Low Exploration}& $0.3$& $0.3$ \\	
 		 		\vspace{0.5cm}		
 		\textbf{Medium Exploration}& $0.5$& $0.5$ \\	
 		 		\vspace{0.5cm}		
 		\textbf{High Exploration}& $0.7$& $0.7$ \\	\hline \hline
	\end{tabular}
 	
 \end{table}
 
 Table \ref{2target} shows that in the first regime with a 2\% inflation target, the steady state velocity is $0.17$. The discount factor is chosen as $0.99$. Exploration level is the standard deviation of the added noise to the policy function (i.e., Equation \ref{OU}). 0.3 represents a low exploration level, 0.5 represents medium exploration, and 0.7 represents high exploration.\footnote{Exploration levels are chosen arbitrarily to reflect their qualitative importance in generating heterogeneous beliefs and behaviours. The quantitative importance of this behavioural parameter is left for future work.}

 Following the simulation protocol I run two experiments:
 
 \begin{itemize}
     \item The first experiment studies the adaptive behaviours of AI agents, and how their different exploration levels impact their beliefs and decisions during an unexpected change in the money supply.\footnote{The exploration parameter has an impact on agents' experience, and it affects agents' learning stages, shown by \cite{Shi2021learning}. This means agents (with different exploration levels) have different beliefs and behaviours in the same aggregate environment facing the identical shock.} To show the impact of exploration solely on agents' transition behaviours during a structural change, I have three AI agents with the same experience but different exploration levels live through the regime change.

     \item The second experiment studies the impact of past experience on agents' beliefs and behaviours during a regime change. To alienate the effect of experience on agents' behaviours, I impose an identical exploration level on all agents.
 \end{itemize}

 

\section{Results and Discussion}
\label{results}

I present the transition dynamics and highlight two main findings: 1. an AI agent that is not expecting a change in the policy target has the ability to adapt its beliefs to this change, and this is reflected by its consumption and demand for real balance decisions. More importantly, AI agents with different levels of exploration behave differently when facing changes in the environment; 2. how quickly the agent response to the changes in its environment depends on its past experience. An agent who has experienced regime shifts will have a smoother transition than an agent who experiences a change for the first time. For all agents, the change happens unexpectedly.


\subsection{Transition Dynamics and Exploration Heterogeneity}

I consider three exploration levels: low, medium and high. Agents with these exploration levels live in the same environment with the same income and interest rate shocks. They also experience the same unforeseen shifts of inflation target. 

\begin{figure}[H]
	\caption{Inflation, three exploration levels}
	\centerline{\includegraphics[width=19cm,height=6cm]{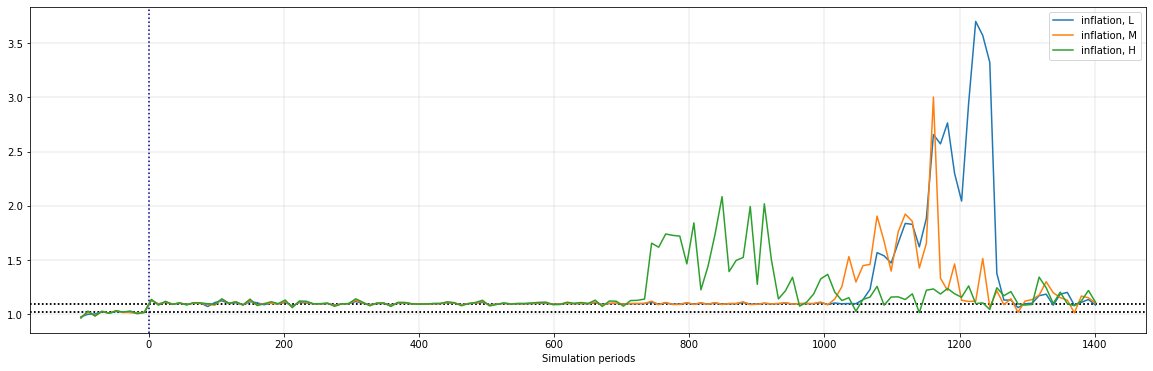}}
	\label{inf}
\end{figure}

\begin{figure}[H]
	\caption{Real balance, three exploration levels}
	\centerline{\includegraphics[width=19cm,height=6cm]{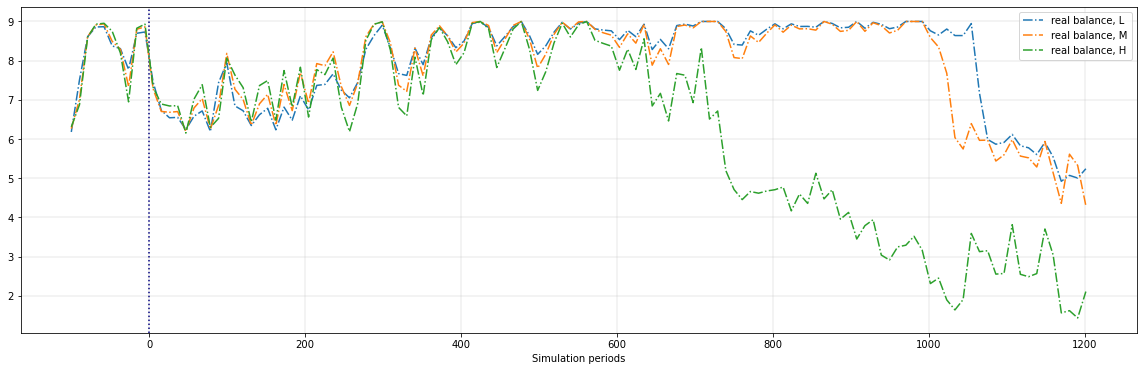}}
	\label{rms}
\end{figure}

\begin{figure}[H]
	\caption{Consumption, three exploration levels}
	\centerline{\includegraphics[width=19cm,height=6cm]{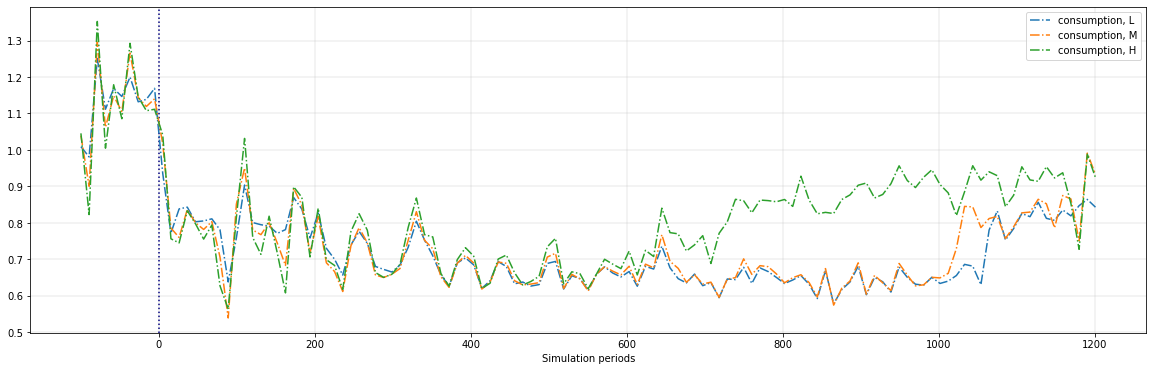}}
	\label{cons}
\end{figure}

\begin{figure}[H]
	\caption{Velocity/ Transaction cost, three exploration levels}
	\centerline{\includegraphics[width=19cm,height=6cm]{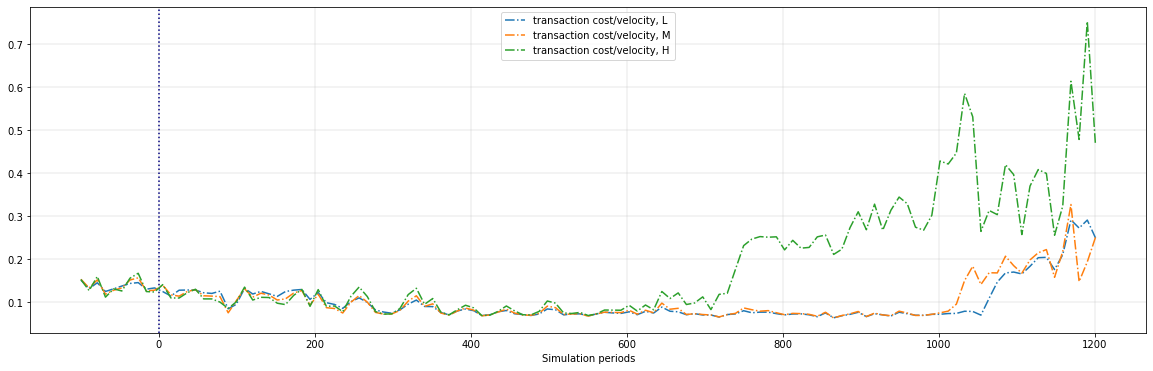}}
	\label{velo}
\end{figure}

Figure \ref{inf} plots the inflation sequence in environments with different exploring AI agents. Period 0 is when the inflation target shifts from 2\% to 10\%. Figure \ref{rms}, \ref{cons} and \ref{velo} plot the real money demand, consumption, and velocity paths of all three agents. The x-axis denotes simulation periods. The vertical black dashed line shows the timing of this target change.

Once the inflation target shifts, figure \ref{rms} shows that agents' real balance demand reduces slightly and then maintains similar levels to their demands before the inflation target change. Money market clearing means that inflation rises to around 10\% (i.e., increased nominal balance and constant real balance) between period 0 and 600. This increase in inflation leads to 40\% reduction in agents' consumption after the regime change (figure \ref{cons}). Lower consumption leads to fewer rewards for these agents. Figure \ref{velo} shows that the transaction cost decreases marginally given the reduced consumption. 

As inflation persists, the agents gather more experience and learn that maintaining a low transaction cost is not beneficial in the high inflation environment with fewer real resources available. A high demand for real balance corresponds to a low transaction cost keeping consumption fixed. They then reduce their real balance demand around period 700, as shown in figure \ref{rms}. Their consumption increases around the same period, as shown in figure \ref{cons}. In relation to the rational expectations steady states, figure \ref{velo} shows that the AI agents shift from close to 0.17 (steady state value of velocity at 2\% inflation) to the region close to 0.32 (steady state value at 10\% inflation).\footnote{This result plots simulations during a regime shift for agents that have learned in the 2\% inflation for periods, hence the velocity value is already close to the steady state value. The learning behaviours to reach this steady state are available upon request.} 

Figure \ref{inf} shows that inflation becomes volatile around period 700, and this is due to learning agents' changes in their demand for real balance. 

Figures \ref{rms}, \ref{cons} and \ref{velo} show that the periods when the agents adjust their behaviours are different, which highlights the importance of the exploration parameter during learning. Figure \ref{cons} shows that at period 600 the high exploration agent (green line) adjusts first, followed by the medium exploration agent. The low exploration agent adjusts last. I observe a similar order of adjustment in figures \ref{rms} and \ref{velo}. This shows that a high exploration agent tends to notice a change in the environment earlier than a low exploration agent. Their differences in adjustment lead to differences in aggregate dynamics. Figure \ref{inf} shows that when the learning agent explores more and adjusts quicker, inflation is less volatile (the green line). Are there any welfare implications from these different speeds of adjustment during a structural change?

\begin{figure}[H]
	\caption{Distribution of rewards, transition periods for three exploration levels}
	\centerline{\includegraphics[width=12cm,height=8cm]{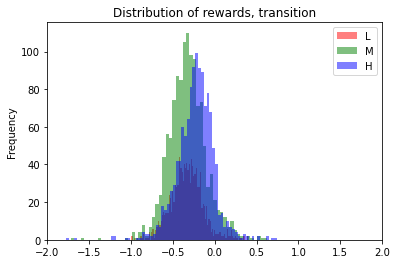}}
	\label{rewards_t}
\end{figure}
\begin{figure}[H]
	\caption{Distribution of rewards, total simulation periods for three exploration levels}
	\centerline{\includegraphics[width=12cm,height=8cm]{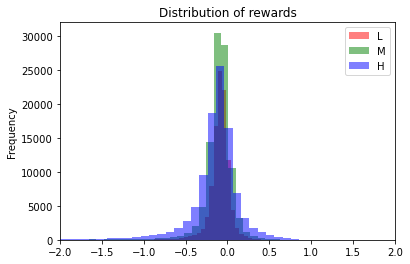}}
	\label{rewards}
\end{figure}

Figures \ref{rewards_t} and \ref{rewards} plot the distribution of rewards for the three types of agents. The blue distribution shows rewards for the high exploration agent. The green distribution shows rewards for the middle exploration agent, and the pink distribution is for the low exploration agent.  

Figure \ref{rewards_t} shows the reward distributions during transition periods. Transition periods are defined as 50 periods before and 1200 periods after the policy regime change. The figure shows that the reward distribution for the high exploration agent has the highest mean, which means that on average the high exploration agent makes decisions that achieve the highest welfare among the three agents. Figure \ref{rewards} plots the reward distributions for the total simulation periods, and it shows that the middle exploration agent's reward distribution has the highest mean. This implies that in an environment that involves changes, undertaking more exploration enables the learning agent to notice and adapt quicker and achieve higher rewards. However, in a stationary environment, this translates into random actions and leads to low overall rewards. 

The result presented in this subsection suggests that with exploration and constant learning, AI agents adjust their actions with respect to a new regime (with a delay). Given the general equilibrium setup of the economic environment, their behaviours impact the aggregate dynamics - that the economy shifts from the neighbourhood of one rational expectation equilibrium to the other. The result shows that, unlike the process in the naive adaptive expectations and the accelerationist controversy literature, AI agents learn to adapt their beliefs with respect to an accelerating money supply. They have different exploration levels, which lead to different learning behaviours and lifetime rewards, and they are less likely to make systematic mistakes.


\subsection{More or Less Experience}
Experience matters to learning, as shown in many empirical studies. I give the learning agent a memory to store all the experience, and belief updating is done by drawing samples from memory. I run simulations for agents with different past experiences. Their exploration levels are identical. 

Three agents are considered: 
\begin{itemize}
    \item agent 1 has never experienced a shift in inflation target; 
    \item agent 2 has experienced the shift once in the past; 
    \item agent 3 has experienced shifts twice in the past. 
\end{itemize}

I compare their simulated paths of consumption during a permanent increase of the inflation target.

  \begin{figure}[H]
	\caption{Consumption Comparison}
	\centerline{\includegraphics[width=19cm,height=8cm]{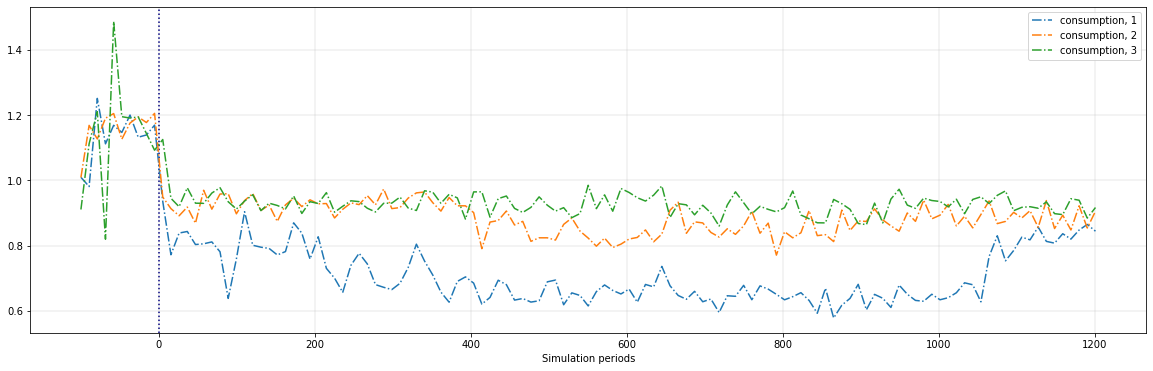}}
	\label{r_mchange}
\end{figure}

Figure \ref{r_mchange} plots consumption paths for the three learning agents with different past experiences. 0 on the x-axis represents the moment when the inflation target changes from 2\% to 10\%. The blue line (consumption 1) denotes the consumption of agent 1, who has never experienced a regime shift. The orange line shows the consumption of agent 2, who has a one time experience of a similar policy shift. The green line denotes agent 3 who has experienced a similar shift twice. 

The agent who has never experienced a regime shift adjusts slower than the other two. Their consumption level goes down when the inflation target increases to 10\%,  and at around period 1000 after this shift, their consumption path starts to adjust to a level that is similar to the other two agents. This lower consumption level translates to lower rewards for this inexperienced agent. 

With these simulations I show that lifetime experience matters for individual behaviours, and is a source of divergence in expectations. This is similar to the empirical evidence presented by \cite{MalmendierNagel2016}. They show that individuals of different ages disagree significantly in their inflation expectations, and this can be explained by differences in their lifetime experiences of inflation. In the simulation experiments here, as an AI agent experiences changes in the monetary policy target, it learns, and reacts more smoothly to a similar shift in the future. Agents who have not experienced a  shift, (e.g., similar to the blue-line agent in Figure \ref{r_mchange}), adjust more slowly and face higher costs (in terms of rewards received).

This experience-based learning depends on the sampling strategies. I draw random samples from agents' memories for all the experiments. Other sampling strategies can be considered, for example, drawing samples from the most recent experiences. It is out of the scope of this paper to discuss which sampling strategy is best; I leave that to the behavioural evidence from the neural science and psychology literature.

\subsection{Discussions}

I present an expectation formation model. I show a transition dynamic for an economy facing a change in monetary policy regime. I also present the mechanism behind agents' adaptive responses to unexpected regime changes.

AI agents under the DRL algorithm becomes aware of policy changes by interacting with their economic environment. This means making a decision given a current state, and observing the next state and the reward signal corresponding to their decision. When the decision-making strategy stops generating a high reward, the agents changes their policy to obtain a higher reward in the long run: this is how an agent adapts their behaviours with respect to a monetary policy target change. 

This is consistent with what \cite{Cavalloetal2017} observe in their experiments. They show that private agents are more likely to adjust their inflation expectations with respect to supermarket price changes than actual inflation statistics. Price changes in supermarkets are more likely to have a direct impact on consumers' perceived welfare than the observed inflation statistics or a central bank announcement on changes to an inflation target. 

One key parameter in this algorithm is exploration. With exploration, agents take random actions, and notice and adapt to structural changes. Another feature is that in a large economy with many learning agents, this parameter can be used to simulate different learning behaviours. This could be an alternative macro simulation lab for policy experiments. 


A prominent criticism of DRL algorithms is the speed of learning. In this paper, it takes a significant number of simulation periods for the agent to adjust their behaviours after a policy shift. This can be accelerated by modifying several training parameters, but the number of simulation periods remains high. As explained by \cite{Botvinick2019}, the slow learning is mainly due to the incremental parameter adjustment and weak inductive bias within the algorithm. However, as it is a fast-evolving literature, many new DRL algorithms are proposed to mitigate this and speed up the learning process. For example, inspired by \cite{Gershman2017}, DRL algorithms with episodic memory are being developed. DRL agents can learn from experience gathered by other agents, and this can also reduce the amount of simulation periods. 



\section{Summary}
 I explore an alternative to the rational expectations hypothesis. I use a deep reinforcement learning algorithm from the artificial intelligence literature to model economic agents’ learning process, and present transition dynamics for an economy with an acceleration in the money supply. The learning agents do not know their own preferences, the underlying economic structure, or how their actions affect the transition dynamics.
 
An AI learning agent born in an unknown environment learns by interacting with the environment. This involves taking exploratory actions and observing stimulus signals (reinforcement learning). The experience is processed by artificial neural networks (deep learning) with the goal of forming a decision-making strategy that maximises the agent's expected (subjective beliefs) future returns. The subjective beliefs evolve based on the agent's experience. 
 
 I simulate an economy to run several experiments. The economy follows a general equilibrium structure with an accelerating money supply. 
 
 With the increase in money supply, the agent (unaware of the change in the money supply) maintains a similar level of real money demand. Money market clearing means that inflation shifts to 10\% (i.e., increased nominal balance and constant real balance). This high inflation takes away real resources, and leads to less consumption, which means fewer rewards for the agent relative to the low inflation scenario. As inflation persists, the agent gathers more experience and learns that maintaining a low transaction cost is not beneficial in the high inflation environment, given the limited real resource. They then reduce their real balance demand and increase consumption to achieve high rewards.
 
 I focus on exploration and its impact on learning and transition behaviours. With a higher level of exploration, the AI agent adjusts more quickly to this shift in inflation target, and obtains higher cumulative rewards during transition periods. A quicker adjustment also leads to a less volatile inflation. However, higher exploration may not be sustainable when the economy is stable over a long period. Randomness leads to low long term rewards in an environment with no shocks. 
 
I show that when an agent has already experienced a change in monetary policy, they respond more quickly and more smoothly to a fresh change, compared to an inexperienced agent. This is similar to empirical evidence presented by \cite{Cavalloetal2017}.

 In future studies, I will simulate economies with heterogeneous learning agents to explore issues such as generational disagreements in inflation expectations and their actions, and cross-country differences in responses to the same macro shocks.
 


\newpage

\bibliographystyle{agsm}
\bibliography{bibtex}

\newpage
\appendix

\end{document}